\documentclass[vecphys,openany]{svmult}

\usepackage{url}
\usepackage{bm}
\usepackage{subfigure}
\usepackage{makeidx}     

\usepackage{multicol}    
\usepackage[bottom]{footmisc}
\usepackage{latexsym,amsfonts,amsmath,amssymb,graphicx}
\makeindex             

\usepackage{epsfig}

\begin{document}

\title*{Beyond The Concept of Manifolds: \\ Principal Trees, Metro
Maps, \\ and Elastic Cubic Complexes}

\toctitle{Beyond The Concept of Manifolds:  Principal Trees, \ \ \
\ \ \ \ \ \ \ \ \ \ \ \ \ \ \ \ \ \ \ \ \ \ \ \ \ \ \ \ \ \ \ \ \
\ \ \ \ \ \ Metro Maps, and Elastic Cubic Complexes}
\titlerunning{Elastic Cubic Complexes and Principal Trees}
\author{Alexander N. Gorban\inst{1,3}\and Neil R. Sumner\inst{1} \and
Andrei Y. Zinovyev\inst{2,3}}

\authorrunning{A.N. Gorban, N.R. Sumner, and
A.Y. Zinovyev}
\institute{University of Leicester, University Road, Leicester,
LE1 7RH,  UK, \\ \texttt{\{ag153,nrs7\}@le.ac.uk} \and Institut
Curie, 26, rue d'Ulm, Paris, 75248, France, \\
\texttt{andrei.zinovyev@curie.fr} \and Institute of Computational
Modeling of Siberian Branch of Russian Academy of Sciences,
Krasnoyarsk, Russia}
%
%
\maketitle

\begin{abstract}
Multidimensional data distributions can have complex topologies
and variable local dimensions. To approximate complex data, we
propose a new type of low-dimensional ``principal object'': a {\it
principal cubic complex}. This complex is a generalization of
linear and non-linear principal manifolds and includes them as a
particular case. To construct such an object, we combine a method
of {\it topological grammars} with the minimization of an elastic
energy defined for its embedment into multidimensional data space.
The whole complex is presented as a system of nodes and springs
and as a product of one-dimensional continua (represented by
graphs), and the grammars describe how these continua transform
during the process of optimal complex construction.

The simplest case of a topological grammar (``add a node'',
``bisect an edge'') is equivalent to the construction of
``principal trees'', an object useful in many practical
applications. We demonstrate how it can be applied to the analysis
of bacterial genomes and for visualization of cDNA microarray data
using the ``metro map'' representation.
\end{abstract}

\keywords{principal trees, topological grammars, principal
manifolds, elastic functional, data visualization}

\section{Introduction and Overview}

In this paper, we discuss a classical problem: how to approximate a
finite set $D$ in $R^m$ for relatively large $m$ by a finite subset
of a regular low-dimensional object in $R^m$. In application, this
finite set is a dataset, and this problem arises in many areas: from
data visualization to fluid dynamics.

The first hypothesis we have to check is: whether the dataset $D$ is
situated near a low--dimensional affine manifold (plane) in $R^m$.
If we look for a point, straight line, plane, ... that minimizes the
average squared distance to the datapoints, we immediately come to
Principal Component Analysis (PCA). PCA is one of the most seminal
inventions in data analysis. Now it is textbook material and
celebrated the 100th anniversary \cite{9Pearson1901}. Nonlinear
generalization of PCA is a great challenge, and many attempts have
been made to answer it. Two of them are especially important for our
consideration: Kohonen's Self-Organizing Maps (SOM) and principal
manifolds.

With the {\it SOM} algorithm \cite{9Kohonen82} we take a finite
metric space $Y$ with metric $\rho$ and try to map it into $R^m$
with (a) the best preservation of initial structure in the image of
$Y$ and (b) the best approximation of the dataset $D$. The  SOM
algorithm has several setup variables to regulate the compromise
between these goals. We start from some initial approximation of the
map, $\phi_1 : Y \rightarrow R^m$. On each ($k$-th) step of the
algorithm we have a datapoint $x \in D$ and a current approximation
$\phi_k : Y \rightarrow R^m$. For these $x$ and $\phi_k$ we define
an ``owner" of $x$ in $Y$: $y_x = {\rm argmin}_{y \in Y} \|
x-\phi_k(y) \|$. The next approximation, $\phi_{k+1}$, is
\begin{equation}\label{SOM}
\phi_{k+1}(y) = h_k w(\rho(y,y_x))( x-\phi_k(y)) \; .
\end{equation}
Here $h_k$ is a step size, $0\leq w(\rho(y,y_x))\leq 1$ is a
monotonically decreasing neighbourhood function. There are many ways
to combine steps (\ref{SOM}) in the whole algorithm. The idea of SOM
is flexible and seminal, it has plenty of applications and
generalizations, but, strictly speaking, we don't know what we are
looking for. We have the algorithm, but no independent definition:
SOM is a result of the algorithm at work. The attempts to define SOM
as solution of a minimization problem for some energy functional
were not very successful \cite{9Erwin92}, however, this led to the
development of the optimization-based Generative Topographic Mapping
(GTM) method \cite{9Bishop1998}.

For a known probability distribution, {\it principal manifolds} were
introduced as lines or surfaces passing through ``the middle'' of
the data distribution \cite{9HastieStuetzle89}. This intuitive
vision was transformed into the mathematical notion of {\it
self-consistency}: every point $x$ of the principal manifold $M$ is
a conditional expectation of all points $z$ that are projected into
$x$. Neither manifold, nor projection need to be linear: just a
differentiable projection $\pi$ of the data space (usually it is
$R^m$ or a domain in $R^m$) onto the manifold $M$ with the
self-consistency requirement for conditional expectations: $
x=\mathbf{E}(z|\pi(z)=x).$ For a finite dataset $D$, only one or
zero datapoints are typically projected into a point of the
principal manifold. In order to avoid overfitting, we have to
introduce smoothers that become an essential part of the principal
manifold construction algorithms.

SOMs give the most popular approximations for principal manifolds:
we can take for $Y$ a fragment of a regular $k$-dimensional grid and
consider the resulting SOM as the approximation to the
$k$-dimensional principal manifold (see, for example,
\cite{9Mulier95,9Ritter92}). Several original algorithms for
construction of principal curves \cite{9Kegl02} and surfaces for
finite datasets were developed during last decade, as well as many
applications of this idea. The recently proposed idea of local
principal curves \cite{lpcLoc} allows to approximate data with
nonlinear, branched, and disconnected one-dimensional continua.

In 1996, in a discussion about SOM at the 5th Russian National
Seminar in Neuroinformatics, a method of multidimensional data
approximation based on elastic energy minimization was proposed (see
\cite{9GorbanRossiev99,9ZinovyevBook00,GorZinComp2005} and the
bibliography there). This method is based on the analogy between the
principal manifold and an elastic membrane (and plate). Following
the metaphor of elasticity, we introduce two quadratic smoothness
penalty terms. This allows one to apply  standard minimization of
quadratic functionals (i.e., solving a system of linear algebraic
equations with a sparse matrix). The elastic map approach led to
many practical applications, in particular in data visualization and
missing data values recovery. It was applied for visualization of
economic and sociological tables
\cite{9GorbanVisPreprint01,9GorbanCHAOS01,9GorbanInfo00,9ZinovyevBook00},
to visualization of natural \cite{9ZinovyevBook00} and genetic texts
\cite{9Gorban03,9GorbanOpSys03}, and to recovering missing values in
geophysical time series \cite{9GapsDGRKM}. Modifications of the
algorithm and various adaptive optimization strategies were proposed
for modeling molecular surfaces and contour extraction in images
\cite{GorZinComp2005}.

\subsection{Elastic Principal Graphs}

Let $G$ be a simple undirected graph with set of vertices $Y$ and
set of edges $E$. For $k \geq 2$ a $k$-star in $G$ is a subgraph
with $k+1$ vertices $y_{0,1, \ldots k} \in Y$ and $k$ edges
$\{(y_0, y_i) \ | \ i=1,\ldots k\} \subset E$. Suppose for each
$k\geq 2$, a family $S_k$ of $k$-stars in $G$ has been selected.
We call a graph $G$ with selected families of $k$-stars $S_k$ an
\index{elastic graph} {\it elastic graph} if, for all $E^{(i)} \in
E $ and $S^{(j)}_k \in S_k$, the correspondent elasticity moduli
$\lambda_i > 0$ and $\mu_{kj}
> 0$ are defined. Let  $E^{(i)}(0),E^{(i)}(1)$ be vertices of an
edge $E^{(i)}$ and $S^{(j)}_k (0),\ldots S^{(j)}_k (k)$ be vertices
of a $k$-star  $S^{(j)}_k $ (among them, $S^{(j)}_k (0)$ is the
central vertex).
 For any map $\phi:Y \to R^m$ the {\it energy of the
graph} is defined as
\begin{eqnarray}
U^{\phi}{(G)}&:=&\sum_{E^{(i)}} \lambda_i
\left\|\phi(E^{(i)}(0))-\phi(E^{(i)}(1)) \right\| ^2 \\ &&+
\sum_{S^{(j)}_k}\mu_{kj} \left\|\sum _ {i=1}^k \phi(S^{(j)}_k
(i))-k\phi(S^{(j)}_k (0)) \right\|^2 \; . \nonumber
\end{eqnarray}

Very recently, a simple but important fact was noticed
\cite{9Gusev04}: every system of elastic finite elements could be
represented by a system of springs, if we allow some springs to have
negative elasticity coefficients.  The energy of a $k$-star $s_k$ in
$R^m$ with $y_0$ in the center and $k$ endpoints $y_{1,\ldots k}$ is
$u_{s_k}= \mu_{s_k}(\sum_{i=1}^k y_i - k y_0)^2$, or, in the spring
representation, $u_{s_k}=k\mu_{s_k} \sum_{i=1}^k (y_i - y_0)^2 -
\mu_{s_k} \sum_{i
> j} (y_i-y_j)^2$. Here we have $k$ positive springs with
coefficients $k\mu_{s_k}$ and $k(k-1)/1$ negative springs with
coefficients $-\mu_{s_k}$.

For a given map $\phi: Y \to R^m$ we divide the dataset $D$ into
subsets $K^y, \, y\in Y$. The set $K^y$ contains the data points for
which the node $\phi(y)$ is the closest one in $\phi(Y)$:

\begin{equation}\label{9taxondef}
K^{y_j} = \{x_i| y_j = \arg \min_{y_k\in Y}\|y_k - x_i\| \}  \; .
\end{equation}

The {\it energy of approximation} is:
\begin{equation}\label{enappr}
U^{\phi}_A(G,D)= \frac{1}{\sum_{ x \in D} w(x)}\sum_{y \in Y}
\sum_{ x \in K^y} w(x) \|x- \phi(y)\|^2 \; ,
\end{equation}
where $w(x) \geq 0$ are the point weights. In the simplest case
$w(x)=1$ but it might be useful to make some points `heavier' or
`lighter' in the initial data. The normalization factor ${1}/{\sum_{
x \in D} w(x)}$ in (\ref{enappr}) is needed for the law of large
numbers\footnote{For more details see Gorban \& Zinovyev paper in
this volume.}.

\section{Optimization of Elastic Graphs \\ for Data Approximation}

\subsection{Elastic Functional Optimization}

The simple algorithm for minimization of the energy
$U^{\phi}=U^{\phi}_A(G,D)+U^{\phi}{(G)}$ is the \index{splitting
algorithm} splitting algorithm, in the spirit of classical $K$-means
clustering:

\vspace*{1mm} \frame{
\begin{minipage}[l]{10.3cm}
\vspace{2mm}
\begin{enumerate}
\item{For a given system of sets $\{K^y \ | \ y \in Y \}$ we
minimize $U^{\phi}$ (it is the minimization of a positive
quadratic functional). This is done by solving a system of linear
algebraic equations for finding new positions of nodes
$\{\phi(y_i)\}$:

\begin{equation}\label{nodeLAE}
\sum_{k = 1}^{p} a_{jk} \phi(y_{k}) = \frac{1}{\sum_{ x \in D}
w(x)}{\sum_{x \in K^{y_j}} w(x) x}\;.
\end{equation}
}

\item{For a given $\phi$ we find new $\{K^y\}$ (\ref{9taxondef}).}

\item{Go to step 1 and so on; stop when there are no significant
changes in $\phi$.}
\end{enumerate}
\vspace{0.01mm}
\end{minipage}
} \vspace{1mm}

\noindent Here,

\begin{equation}\label{9matrix}
 a_{jk} = \frac{n_j \delta_{jk}} {\sum_{ x
\in D} w(x)} + e_{jk}  + s_{jk} ,  \;\;\; n_j = \sum_{x \in
K^{y_j}}w(x) \;\;\; (j = 1\ldots p) \; ,
\end{equation}
$\delta_{jk}$ is Kronecker's $\delta$, and matrices $e_{jk}$ and
$s_{jk}$ depend only on elasticity modules and on the content of
the sets $\{E^{(i)}\}$ and $\{S^{(i)}_k\}$, thus they need not be
recomputed if the structure of the graph was not changed.

Matrix $a_{jk}$ is sparse. In practical computations it is easier
to compute only non-zero entries of the $e_{jk}$ and $s_{jk}$
matrices. This can be done using the following scheme:

\vspace*{1mm} \frame{
\begin{minipage}[l]{10.3cm}
\vspace{2mm}
\begin{enumerate}
\item{Initialize the $s_{ij}$ matrix to zero.}
\item{For each $k$-star $S^{(i)}_k$ with weight $\mu_i$, outer nodes
$y^{N_1},\ldots,y^{N_k}$ and central node $y^{N_0}$, the $s_{ij}$
matrix is updated as follows ($1\leq l,m\leq k$):
$$\begin{array}{ll} s_{N_0N_0}'=s_{N_0N_0}+k^2\mu_i,\;\;
s_{N_lN_m}'=s_{N_lN_m}+\mu_i \; ,\\
s_{N_0N_l}'=s_{N_0N_l}-k\mu_i,\;\; s_{N_lN_0}'=s_{N_lN_0}-k\mu_i
\; .
\end{array}$$}
\item{Initialize the $e_{ij}$ matrix to zero.}
\item{For each edge $E^{(i)}$ with weight \textit{$\lambda
$}$_{i}$, one vertex $y^{k1}$ and the other vertex $y^{k2}$, the
$e_{jk}$ matrix is updated as follows: $$\begin{array}{ll}
e_{k_{1} k_{1}}  = e_{k_{1} k_{1}}  + \lambda _{i} \;  ,\;\;
e_{k_{2} k_{2}} = e_{k_{2} k_{2}}  + \lambda _{i}  \; , \\
e_{k_{1} k_{2}} = e_{k_{1} k_{2}}  - \lambda _{i} \;  , \;\;
e_{k_{2} k_{1}} = e_{k_{2} k_{1}}  - \lambda _{i}\;. \end{array}
$$}
\end{enumerate}
\vspace{0.01mm}
\end{minipage}
}

\vspace*{1mm}

This algorithm gives a local minimum, and the global minimization
problem arises. There are many methods for improving the
situation, but without guarantee of finding the global
minimization (see, for example, accompanying paper
\cite{GorZin2007Springer}).

\subsection{Optimal Application of Graph Grammars}

The next problem is the elastic graph construction. Here we should
find a compromise between simplicity of graph topology, simplicity
of geometrical form for a given topology, and accuracy of
approximation. Geometrical complexity is measured by the graph
energy $U^{\phi}{(G)}$, and the error of approximation is measured
by the energy of approximation $U^{\phi}_A(G,D)$. Both are
included in the energy $U^{\phi}$. Topological complexity will be
represented by means of elementary transformations: it is the
length of the energetically optimal chain of elementary
transformation from a given set applied to the initial simple
graph.

The graph grammars \cite{Nagl,Loewe} provide a well-developed
formalism for the description of elementary transformations. An
\index{graph grammar} elastic graph grammar is presented as a set
of \index{production rule} production (or substitution) rules.
Each rule has a form $A \to B$, where $A$ and $B$ are elastic
graphs. When this rule is applied to an elastic graph, a copy of
$A$ is removed from the graph together with all its incident edges
and is replaced with a copy of $B$ with edges that connect $B$ to
the graph. For a full description of this language we need the
notion of a {\it labeled graph}. Labels are necessary to provide
the proper connection between $B$ and the graph.

A link in the energetically optimal transformation \index{optimal
transformation} chain is constructed by finding a transformation
application that gives the largest energy descent (after an
optimization step), then the next link, and so on, until we
achieve the desirable accuracy of approximation, or the limit
number of transformations (some other termination criteria are
also possible). The selection of an energetically optimal
application of transformations by the trial optimization steps is
time-consuming. There exist alternative approaches. The
preselection of applications for a production rule $A \to B$ can
be done through the comparison of the energy of copies of $A$ with
its incident edges and stars in the transformed graph $G$.

\subsection{Factorization and Transformation of Factors}

\begin{figure}[t]
\centering{\includegraphics[width=80mm, height=30mm]{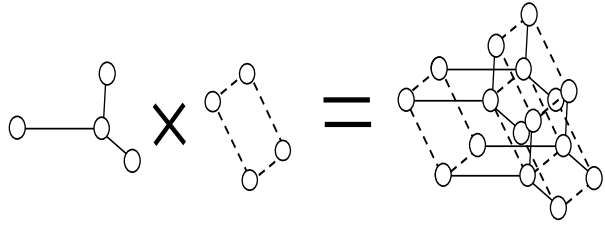} }
\caption{Cartesian product of graphs \label{FigFactor}}
\end{figure}

If we approximate multidimensional data by a $k$-dimensional
object, the number of points (or, more generally, elements) in
this object grows with $k$ exponentially. This is an obstacle for
grammar--based algorithms even for modest $k$, because for
\index{factorization} analysis of the rule $A \to B$ applications
we should investigate all isomorphic copies of $A$ in $G$. The
natural way to avoid this obstacle is the principal object
factorization. Let us represent an elastic graph as a Cartesian
product of graphs (Fig.~\ref{FigFactor}).

The Cartesian product $G_1 \times \ldots \times G_r$ of elastic
graphs $G_1, \ldots G_r$ is an elastic graph with vertex set $V_1
\times \ldots \times V_r$. Let $1 \leq i \leq r$ and $v_j \in V_j$
($j \neq i$). For this set of vertices, $\{v_j\}_{j\neq i}$, a
copy of $G_i$ in $G_1 \times \ldots \times G_r$ is defined with
vertices $(v_1,\ldots v_{i-1}, v, v_{i+1}, \ldots v_r)$ ($v\in
V_i$), edges $$((v_1,\ldots v_{i-1}, v, v_{i+1}, \ldots v_r),
(v_1,\ldots v_{i-1}, v', v_{i+1}, \ldots v_r)), \;\; (v,v') \in
E_i \; ,$$  and, similarly, $k$-stars of the form $(v_1, \ldots
v_{i-1}, S_k, v_{i+1},\ldots v_r)$, where $S_k$ is a $k$-star in
$G_i$. For any $G_i$ there are $\prod_{j, j\neq i} |V_j|$ copies
of $G_i$ in $G$. Sets of edges and $k$-stars for Cartesian product
are unions of that set through all copies of all factors. A map
$\phi: V_1 \times \ldots \times V_r \to R^m$ maps all the copies
of factors into $R^m$ too. {\it The energy of the elastic graph
product is the energy sum of all factor copies.} It is, of course,
a quadratic functional of $\phi$.

The only difference between the construction of general elastic
graphs and factorized graphs is in the application of the
transformations. For factorized graphs, we apply them to factors.
This approach significantly reduces the amount of trials in
selection of the optimal application. The simple grammar with two
rules, ``add a node to a node, or bisect an edge," is also
powerful here, it produces products of primitive elastic trees.
For such a product, the elastic structure is defined by the
topology of the factors.

\section{Principal Trees (Branching Principal Curves)}

\subsection{Simple Graph Grammar (``Add a Node'', ``Bisect an Edge'')}

As a simple (but already rather powerful) example  we use a system
of two transformations: ``add a node to a node" and ``bisect an
edge." These transformations act on a class of {\it primitive
elastic graphs}:  all non-terminal nodes with $k$ edges are
centers of elastic k-stars, which form all the $k$-stars of the
graph. For a primitive elastic graph, the number of stars is equal
to the number of non-terminal nodes -- the graph topology
prescribes the elastic structure. \index{principal tree}

\begin{figure}[t]
\centering{
\includegraphics[width=50mm]{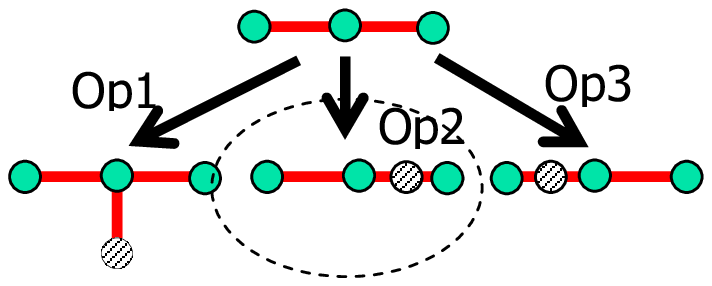}\hspace{10mm}
\includegraphics[width=50mm]{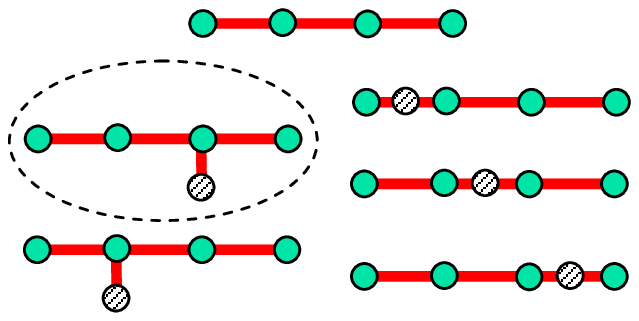}\\
\textbf{a}) \hspace{50mm} \textbf{b}) \\
\includegraphics[width=50mm]{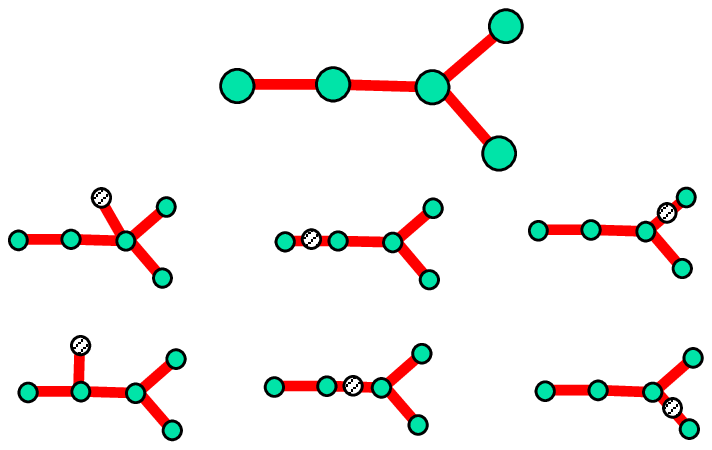}\\
\textbf{c}) } \caption{\label{transformations}Illustration of the
simple ``add node to a node'' or ``bisect an edge'' graph grammar
application. \textbf{a}) We start with a simple 2-star from which
one can generate three distinct graphs shown. The ``Op1''
operation is adding a node to a node, operations ``Op1'' and
``Op2'' are edge bisections (here they are topologically
equivalent to adding a node to a terminal node of the initial
2-star). For illustration let us suppose that the ``Op2''
operation gives the biggest elastic energy decrement, thus it is
the ``optimal'' operation. \textbf{b}) From the graph obtained one
can generate 5 distinct graphs and choose the optimal one.
\textbf{c}) The process is continued until a definite number of
nodes is inserted}
\end{figure}

\begin{figure}[t]
\centering{
\includegraphics[width=37mm, height=35mm]{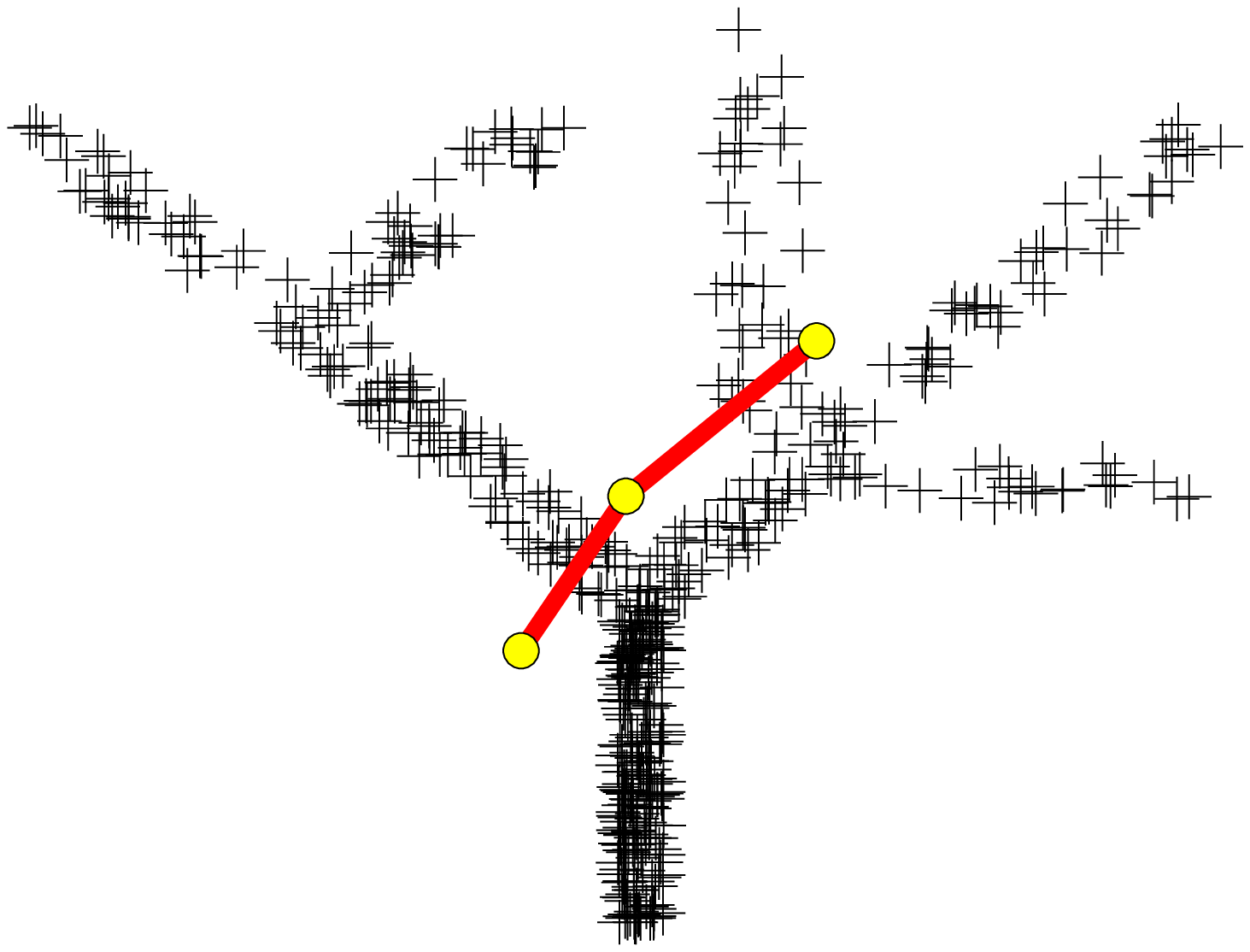}
\includegraphics[width=37mm, height=35mm]{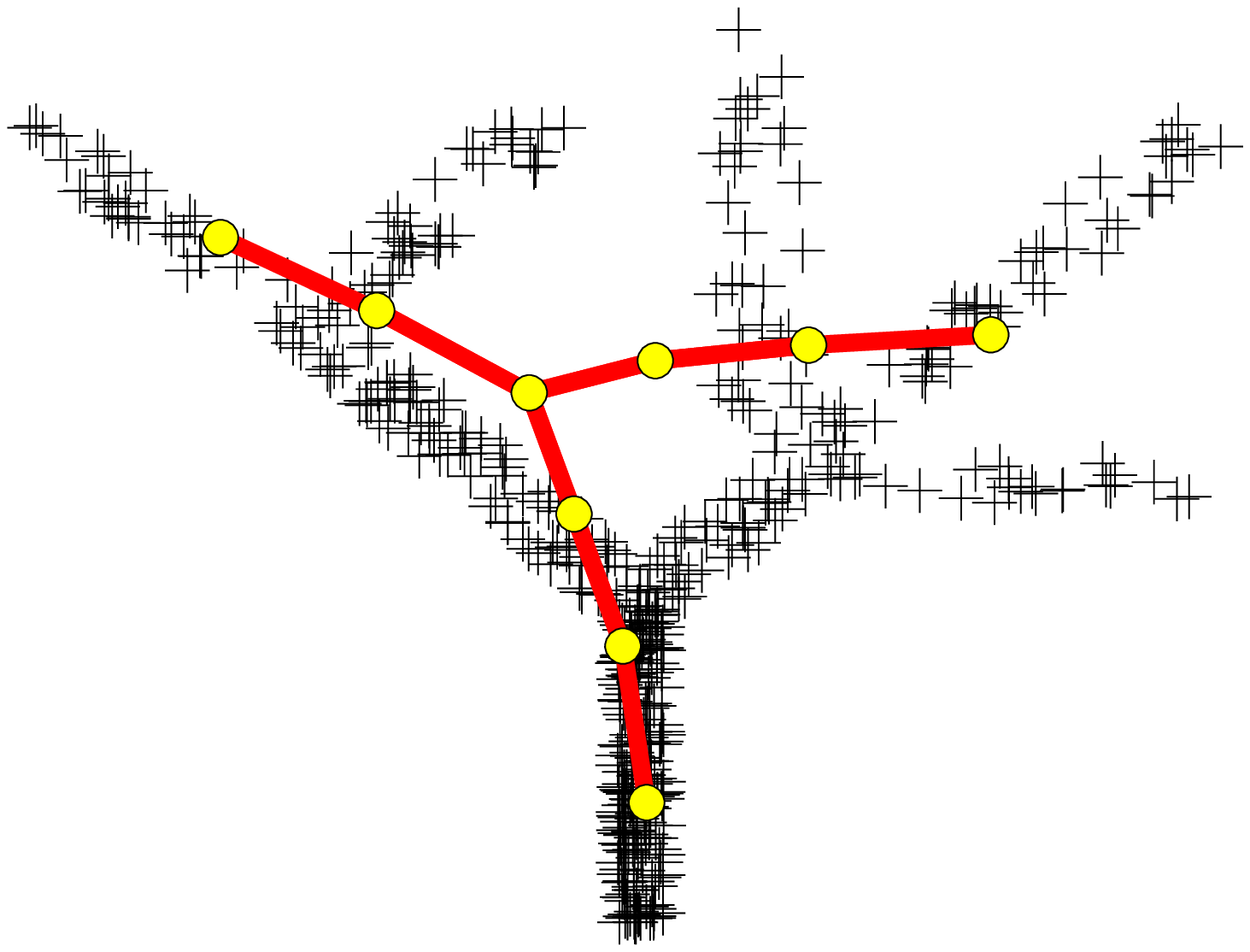}
\includegraphics[width=37mm, height=35mm]{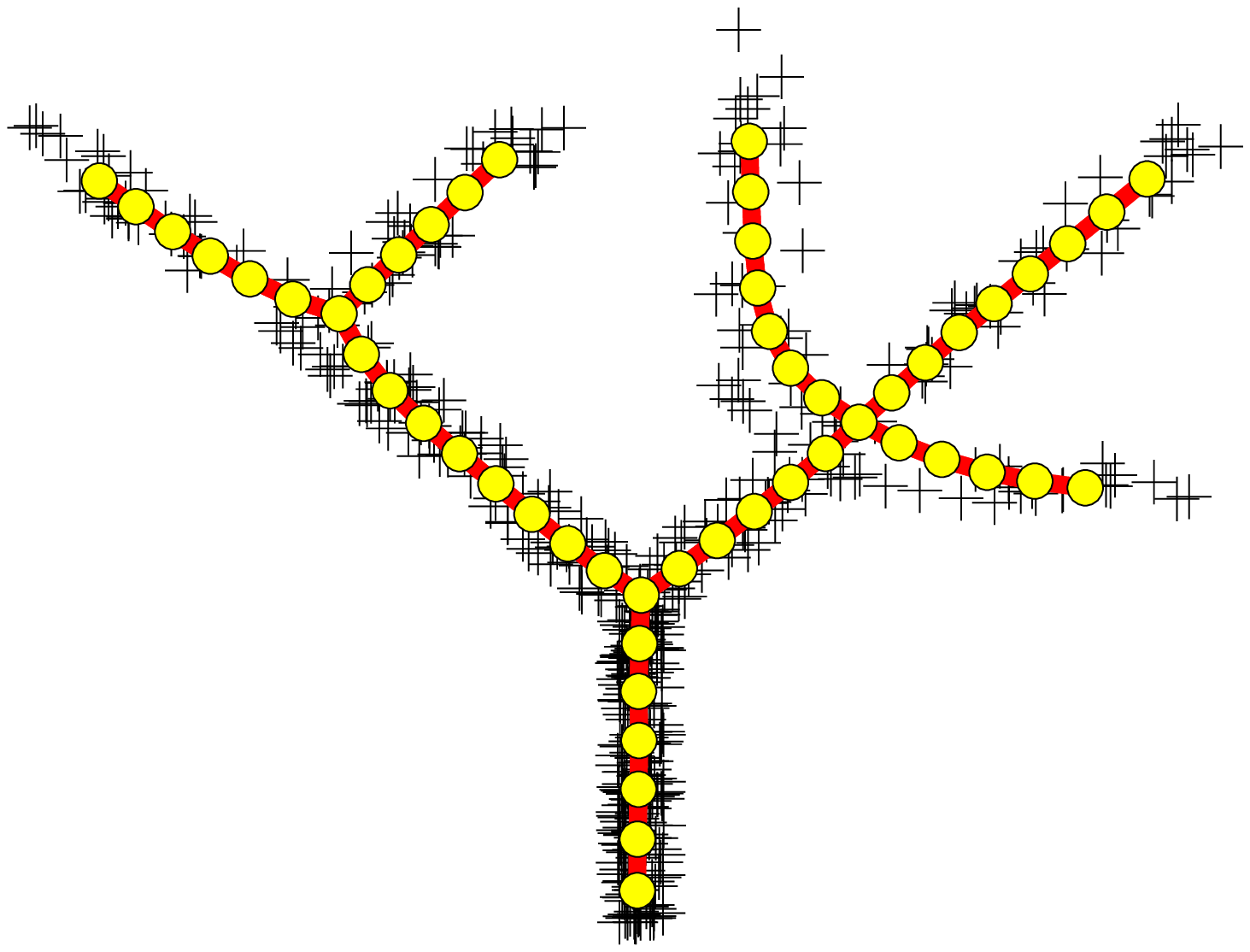}
\\
Iteration 1 \hspace{2.2cm} Iteration 5 \hspace{2.2cm} Iteration 50
\\
\includegraphics[width=37mm, height=35mm]{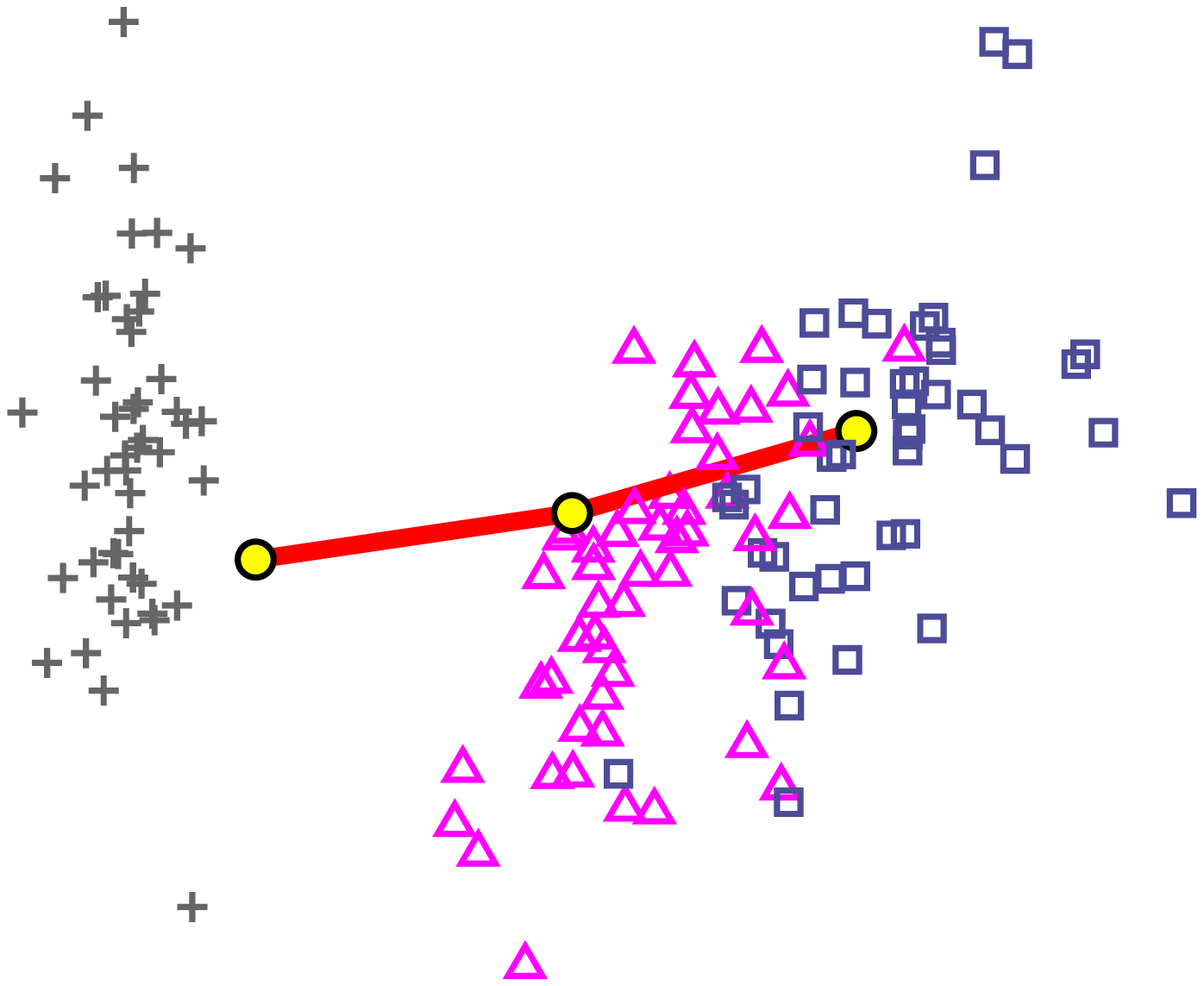}
\includegraphics[width=37mm, height=35mm]{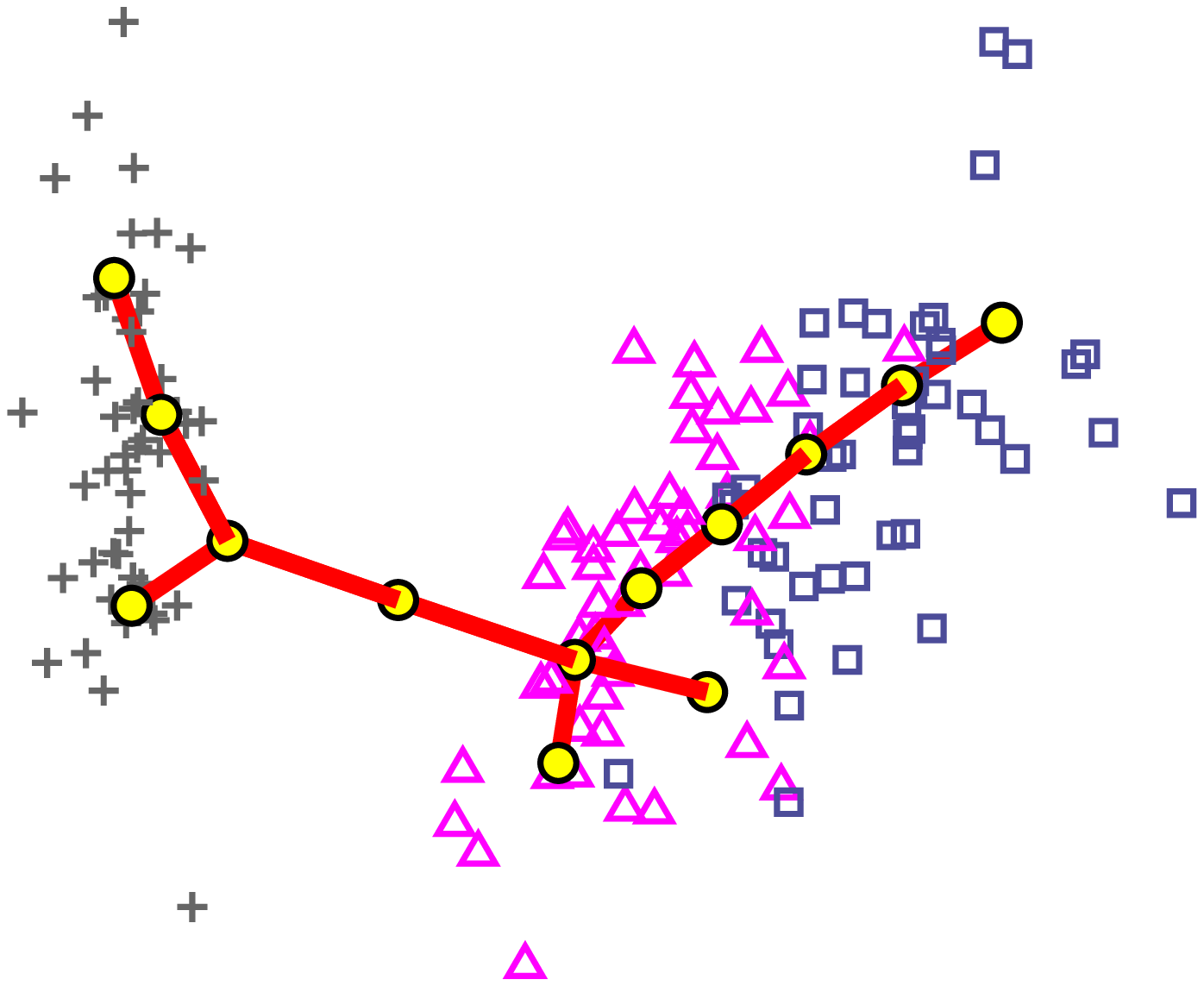}
\includegraphics[width=37mm, height=35mm]{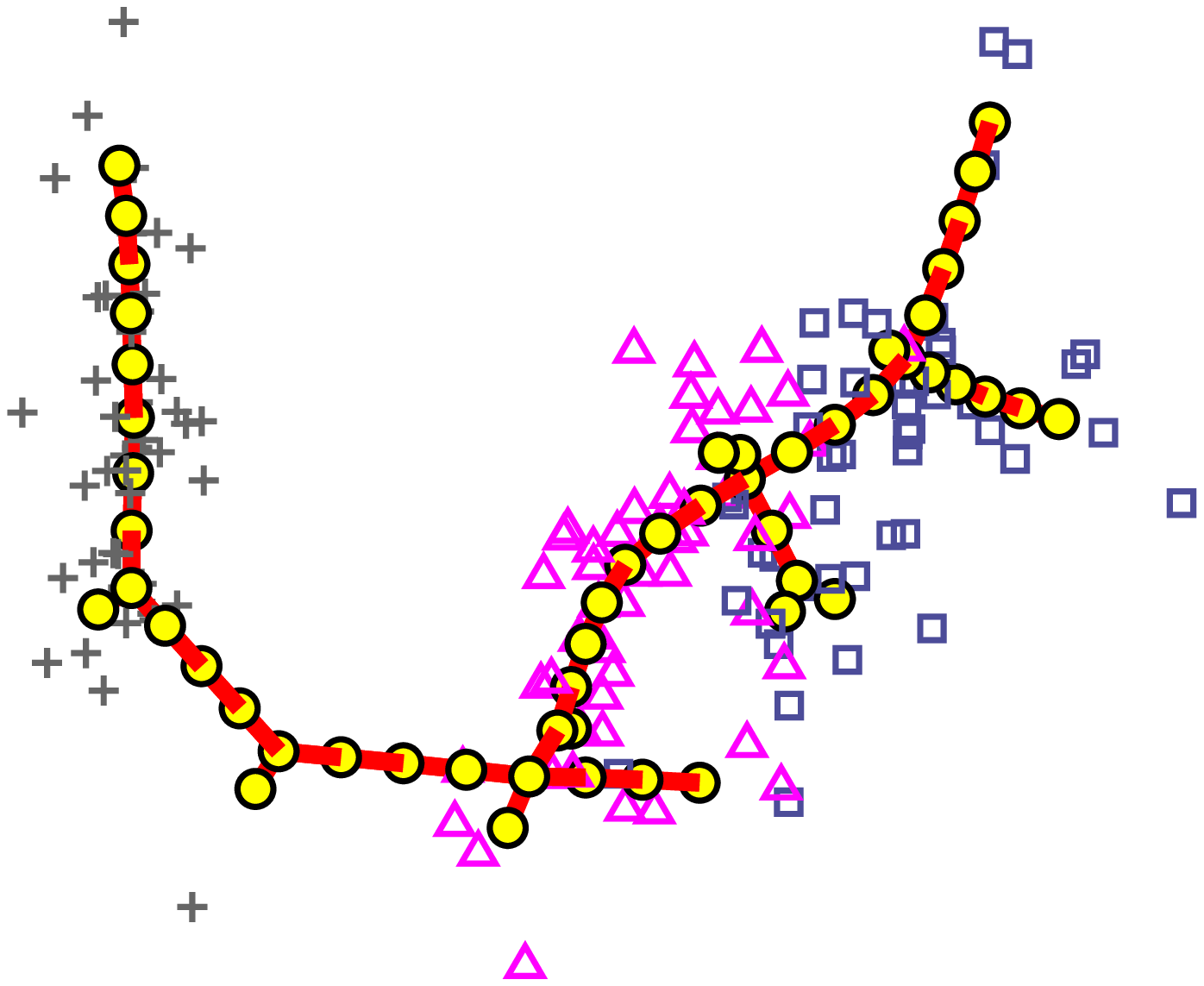}
\\
Iteration 1 \hspace{2.2cm} Iteration 10 \hspace{2.2cm} Iteration
50 } \caption{\label{examples}Applying a simple ``add a node to a
node or bisect an edge" grammar to construct principal elastic
trees (one node is added per iteration). Upper row: an example of
two-dimensional branching distribution of points. Lower row: the
classical benchmark, the  ``iris" four-dimensional dataset (point
shapes distinguish three classes of points), the dataset and
principal tree are presented in projection onto the plane of the
first two principal components}
\end{figure}

The transformation {\it ``add a node"} can be applied to any
vertex $y$ of $G$:  add a new node $z$ and a new edge $(y,z)$. The
transformation {\it ``bisect an edge"} is applicable to any pair
of graph vertices $y,y'$ connected by an edge $(y,y')$: Delete
edge $(y,y')$, add a vertex $z$ and two edges, $(y,z)$ and
$(z,y')$. The transformation of the elastic structure (change in
the star list) is induced by the change of topology, because the
elastic graph is primitive. This two--transformation grammar with
energy minimization builds {\it principal trees} (and principal
curves, as a particular case) for datasets. A couple of examples
are presented on Fig.~\ref{examples}. For applications, it is
useful to associate one-dimensional continua with these principal
trees. Such a continuum consists of node images $\phi(y)$ and of
pieces of straight lines that connect images of linked nodes.

\subsection{Visualization of Data Using \\ ``Metro Map'' Two-Dimensional Tree Layout}

A principal tree is embedded into a multidimensional data space.
It approximates the data so that one can project points from the
multidimensional space into the closest node of the tree (other
projectors are also possible, for example, see the accompanying
paper \cite{GorZin2007Springer}). The tree by its construction is
a one-dimensional object, so this projection performs dimension
reduction of the multidimensional data. The question is how to
represent the result of this projection? For example, how to
produce a tree layout on the two-dimensional surface of paper
sheet? Of course, there are many ways to layout a tree on a plane.
It is always possible to find a tree layout without edge
intersection. But it would be very nice if both some tree
properties and global distance relations would be represented
using the layout. We can require that

1) In a two-dimensional layout, all k-stars should be represented
equiangular, because the ideal configuration of a k-star with
small energy is equiangular and with equal edge lengths.

2) The edge lengths should be proportional to their length in the
multi-dimensional embedding; thus one can represent between-node
distances.
\begin{figure}[t]
\centerline{\includegraphics[width=8cm]{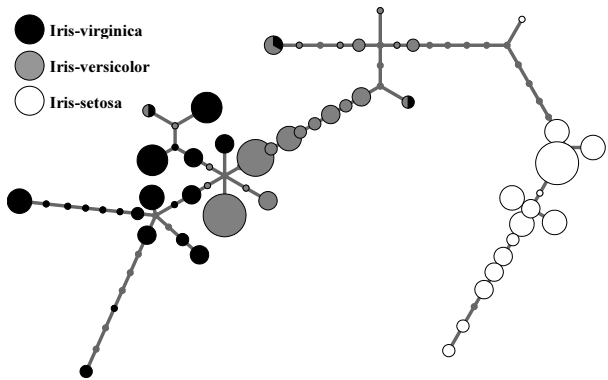}}
\caption{Two-dimensional layout of the principal tree constructed
for the iris dataset. In this layout all stars are equiangular and
the edge lengths are proportional to the real edge lengths in the
multidimensional space. The data points were projected into the
closest tree nodes. The circle radii are proportional to the number
of points projected into each node. Three different colors (or gray
tints on the gray version of the image) denote three
 point classes. If points of different classes were
projected into the same tree node then the number of points of
every class is visualized by the pie diagram}
\label{iristreelayout}
\end{figure}

This defines a tree layout up to global rotation and scaling and
also up to changing the order of leaves in every $k$-star. We can
change this order to eliminate edge intersections, but the result
can not be guaranteed. In order to represent the global distance
structure, we found that a good approximation for the order of
$k$-star leaves can be taken from the projection of every $k$-star
on the linear principal plane calculated for all data points, or
on the local principal plane in the vicinity of the $k$-star,
calculated only for the points close to this star.

In the current implementation of the method, after defining the
initial approximation, a user manually modifies the layout
switching the order of $k$-star leaves to get rid of edge
intersections. The edge lengths are also modifiable. Usually it is
possible to avoid edge intersections in the layout after a small
number of initial layout modifications. This process could be also
fully automated, by a greedy optimization algorithm, for example,
but this possibility is not yet implemented.

The point projections are then represented as pie diagrams, where
the size of the diagram reflects the number of points projected
into the corresponding tree node. The sectors of the diagram allow
us to show proportions of points of different classes projected
into the node (see an example on Fig.~\ref{iristreelayout}).

We call this type of visualization a ``metro map'' since it is a
schematic and ``idealized'' representation of the tree and the data
distribution with inevitable distortions made to produce a nice 2D
layout, but using this map one can still estimate the distance from
a point (tree node) to a point passing through other points. This
map is inherently unrooted (as a real metro map). It is useful to
compare this metaphor with trees produced by hierarchical clustering
where the metaphor is closer to a ``genealogy tree''.

\subsection{Example of Principal Cubic Complex: \\
Product of Principal Trees}

Principal trees are one-dimensional objects. To illustrate the idea
of a $d$-dimensional principal cubic complex that is a cartesian
product of graphs (see Fig.~\ref{FigFactor}), we implemented an
algorithm, constructing products of principal trees. The pseudocode
for this algorithm is provided below:

\vspace*{1mm} \frame{
\begin{minipage}[l]{10.3cm}
\vspace{2mm}
\begin{enumerate}
\item{Initialize one factor graph consisting of one edge
connecting two nodes positioned half a standard deviation around the
data mean.}
\item{Optimize the node positions.}
\item{Test the addition of a node to each star in each factor,
optimizing the node postions of the Cartesian product.}
\item{Test the bisection of each edge in each factor, optimising the
node positions of the Cartesian product.}
\item{ Test the initialisation of another factor graph consisting of
one edge using the scheme:-
    \begin{enumerate}
    \item{ For each node use the k-means algorithm with k=2 on the data class associated with the node, intializing using the mean of the data class and the node
    position.}
    \item{Normalize the vector between the 2 centres thus obtained and scale it to the size of the mean of the edge lengths already incident with the
    node.}
    \item{ Optimize the node positions.}
    \end{enumerate}}
\item{ Choose the transformation from steps (3) to (5) which gives the greatest energy
descent.}
\item{Until the stopping criteria are met repeat steps (3) - (6).}
\end{enumerate}
\vspace{0.01mm}
\end{minipage}
} \vspace{1mm}

\begin{figure}[t]
\begin{center}
\includegraphics[width=6cm,height=6cm]{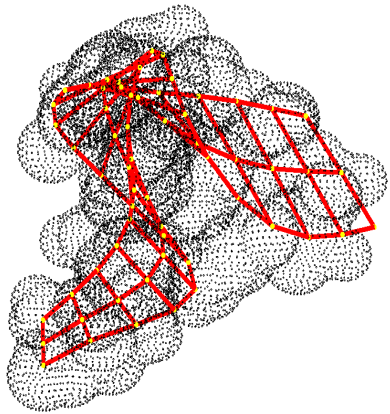}
\end{center}
\caption{A cubic complex (2-dimensional, ``add a node, bisect an
edge'' graph grammar) constructed for a distribution of points on
the Van-der-Waals surface of a fragment of a DNA molecule (dots).
The result is the product of two unbranched trees which
``discovers'' the double-helical DNA structure \label{dooda}}
\end{figure}

 Thus, the structure of a principal cubic complex is defined by its
dimension and the graph grammar applied for its construction. A
simple example of 2-dimensional principal tree, i.e. cubic complex
constructed with the simplest ``add a node or bisect an edge''
grammar (see Fig.~\ref{transformations}) is given in
Fig.~\ref{dooda}. Here the cubic complex is constructed for a
distribution of points on the molecular surface of a fragment of a
DNA molecule (compare with the application of the method of elastic
maps to the same dataset, given in the accompanying paper
\cite{GorZin2007Springer}). The method of topological grammars
\cite{9Gorban07AML} ``discovers'' the double-helical structure of
the DNA molecule. Note that in this example the energy optimization
gave no branching in both factors (trees) and as a result we
obtained a product of two simple poly-lines. In other situations,
the resulting cubic complex could be more complicated, with
branching in one or both factors.

\section{Analysis of the Universal 7-Cluster Structure \\ of Bacterial Genomes}

In this section we describe the application of the method of
topological grammars to the analysis of the cluster structure of
bacterial genomes. This structure appears as a result of projecting
a genome sequence into a multidimensional space of short word
frequencies \cite{7cluPhA,7clulast}. In particular, we show that a
one-dimensional principal tree can reveal a signal invisible when
three-dimensional PCA is applied.

\subsection{Brief Introduction}

One of the most exciting problems in modern science is to
understand the organization of living matter by reading genomic
sequences. The information that is needed for a living cell to
function is encoded in a long molecule of DNA. It can be presented
as a text that has only four letters A, C, G and T.

\index{genome}One distinctive message in a genomic sequence is a
piece of text, called a {\it gene}. Genes can be oriented in the
sequence in the forward and backward directions. In bacterial
genomes genes are always continuous from their start to the stop
signal.

\index{codon}It was one of many great discoveries of the twentieth
century that biological information is encoded in genes by means of
triplets of letters, called {\it codons} in the biological
literature. In the famous paper by Crick {\it et al}.
\cite{Crick61}, this fact was proven by genetic experiments carried
out on bacteria mutants.

In nature, there is a special mechanism that is designed to read
genes. It is evident that as the information is encoded by
non-overlapping triplets, it is critical for this mechanism to
start reading a gene without a shift, from the first letter of the
first codon to the last one; otherwise, the information decoded
will be completely corrupted.

A {\it word} is any continuous piece of text that contains several
subsequent letters. As there are no spaces in the text, separation
into words is not unique.

The method we use to ``decipher'' genomic sequences is the
following. We clip the whole text into fragments of 300 letters in
length and calculate the frequencies of short words (of length 1--4)
inside every fragment. This gives a description of the text in the
form of a numerical table (word frequency vs fragment number).

As there are only four letters, there are four possible words of
length $1$ (singlets), $16=4^2$ possible words of length $2$
(duplets), $64=4^3$ possible words of length $3$ (triplets) and
$256=4^4$ possible words of length $4$ (quadruplets). The first
table contains four columns (frequency of every singlet) and the
number of rows equals the number of fragments. The second table has
16 columns and the same number of rows, and so on.

\begin{figure}[t]
\centering{
\begin{tabular}{lr}

\textbf{a}) \includegraphics[width=50mm, height=50mm]{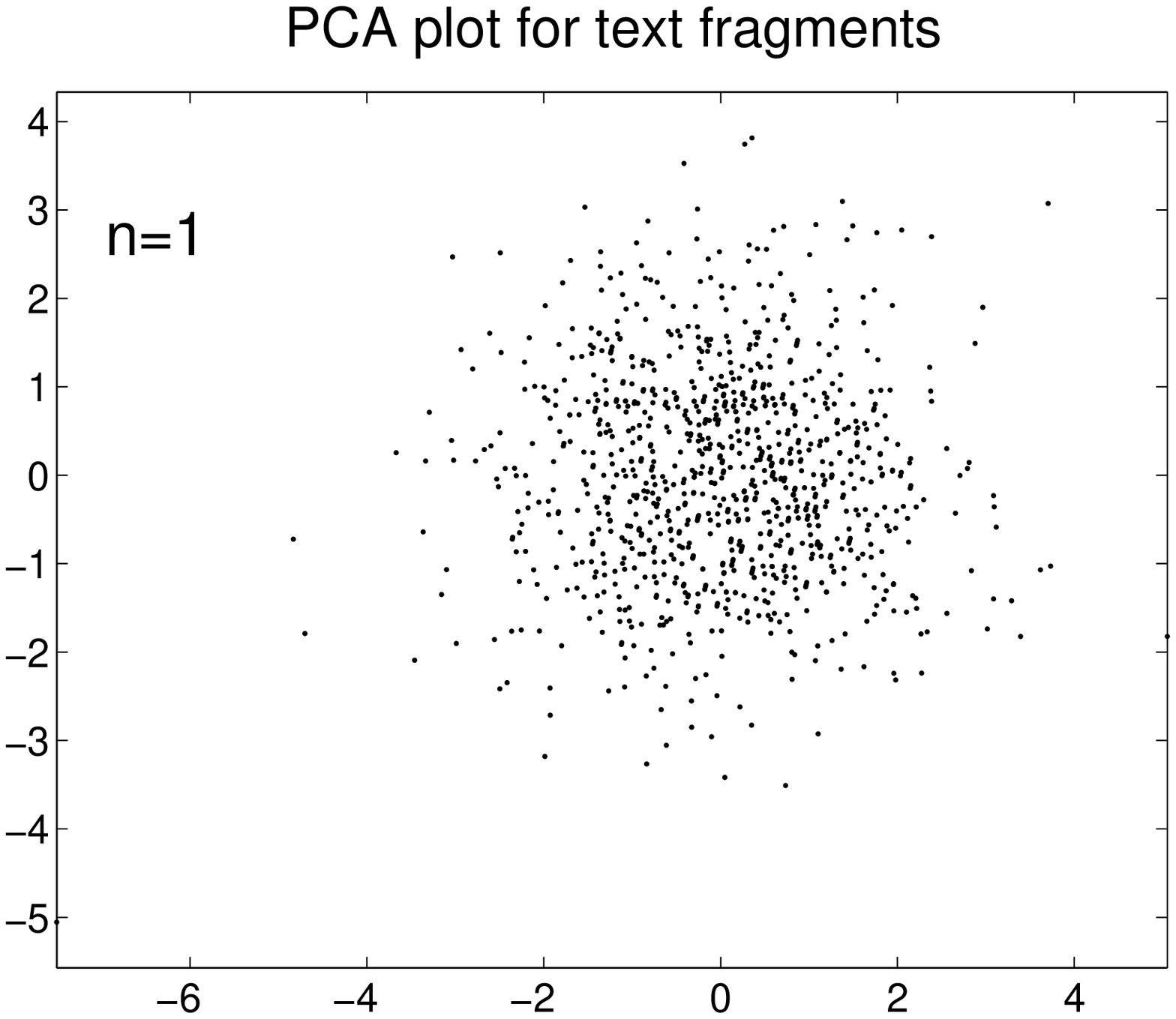}
 &
\textbf{b}) \includegraphics[width=50mm, height=50mm]{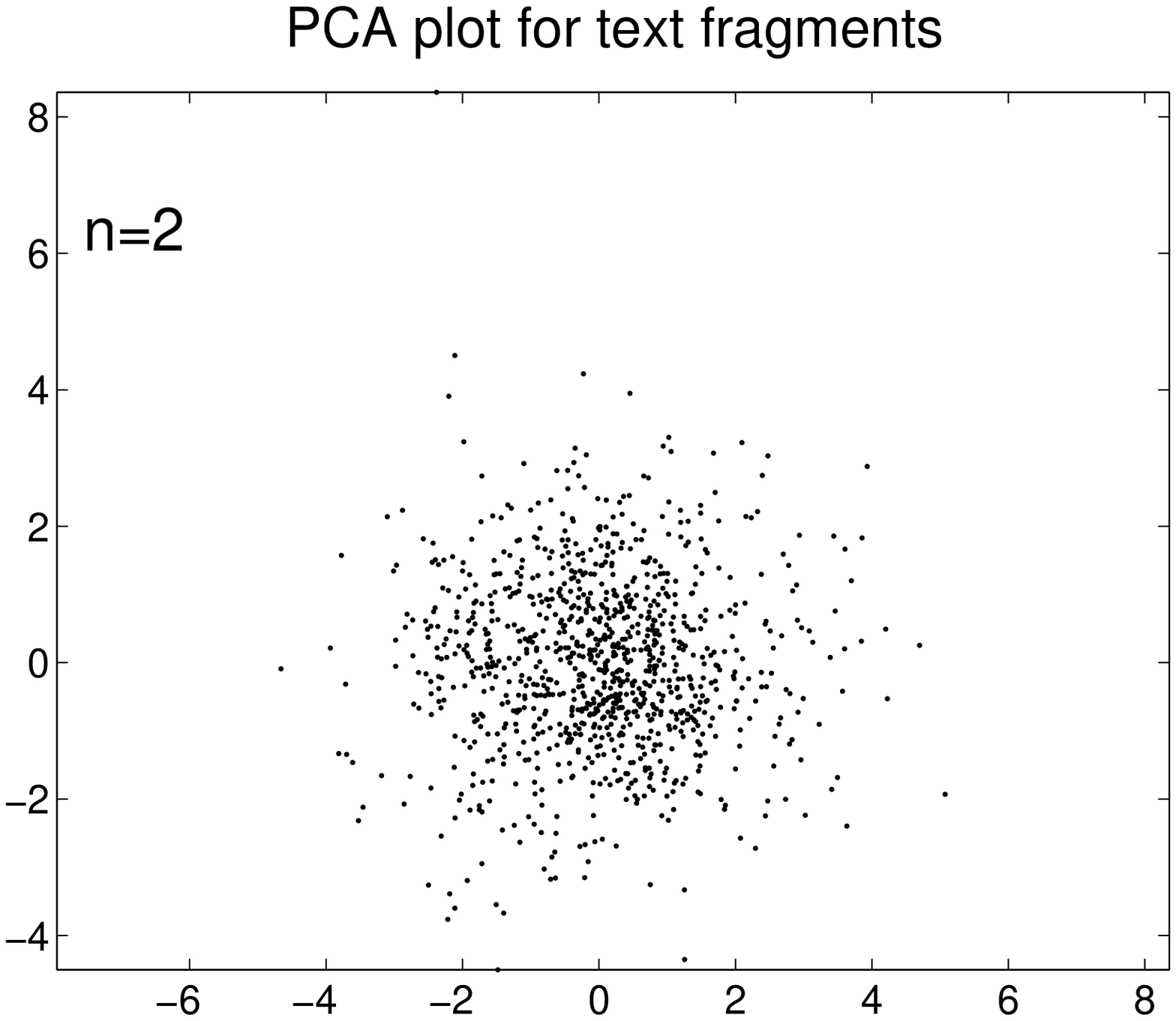}
  \\
\textbf{c}) \includegraphics[width=50mm, height=50mm]{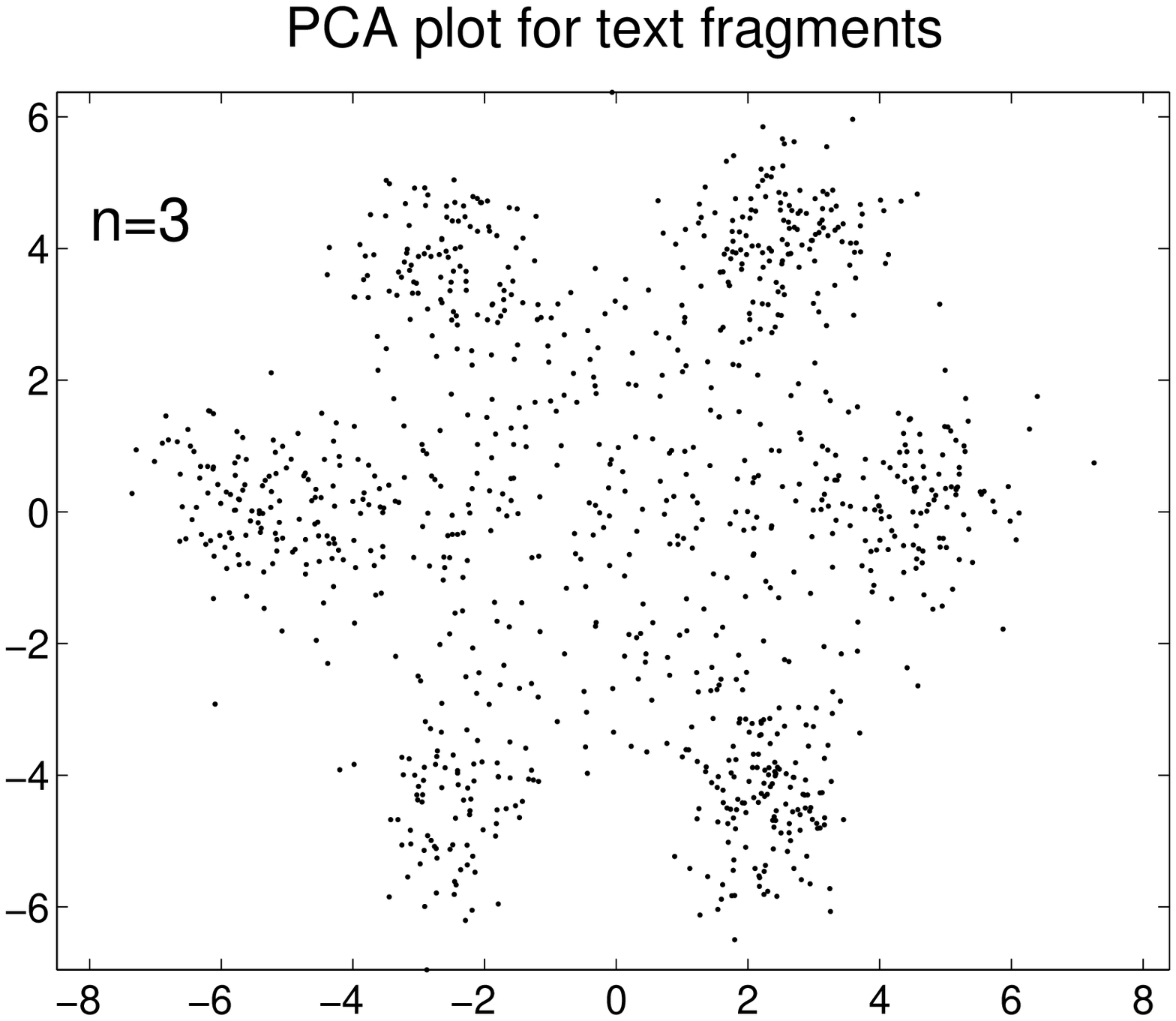}
 &
\textbf{d}) \includegraphics[width=50mm, height=50mm]{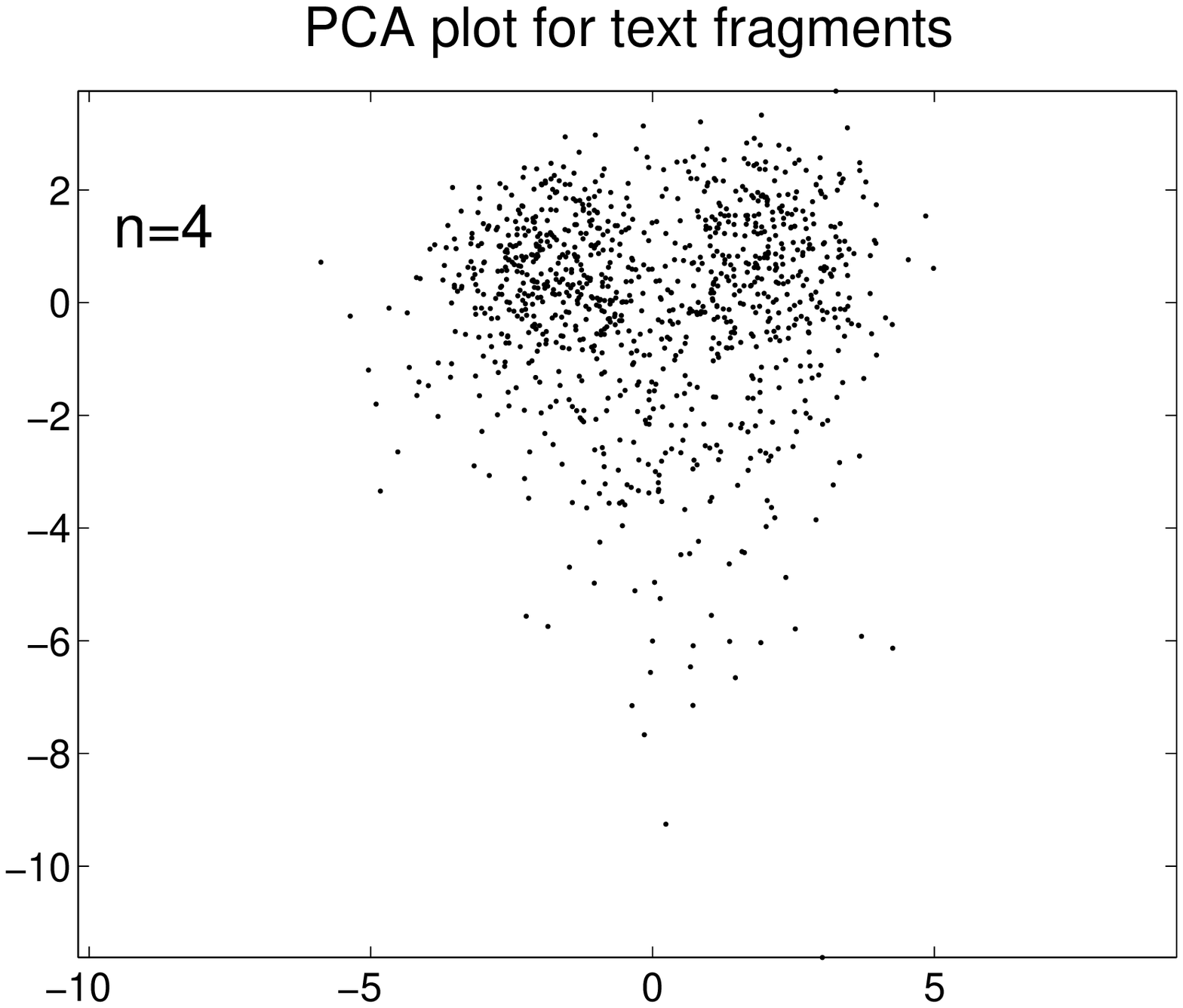}

\end{tabular}
}\caption{PCA plots of word frequencies of different length. In
\textbf{c}) one can see the most structured distribution. The
structure is interpreted as the existence of a non-overlapping
triplet code \label{pcas}}
\end{figure}

These tables can be visualized by means of standard PCA. The result
of such visualization is given on Fig.~\ref{pcas}. As one can see
from PCA plots, counting triplets gives an interesting flower-like
pattern (described in details in \cite{7cluPhA,7clulast}), which can
be interpreted as the existence of non-overlapping triplet code.

The triplet picture evidently contains 7 clusters, and it is more
structured in the space than 1,2- and 4-tuples. To understand the
7-cluster structure, let us make some explanations.

Let us blindly cut the text into fragments. Any fragment can
contain: (a) piece of gene in the forward direction; (b) piece of
gene in the backward direction; (c) no genes (non-coding part);
(d) a mixture of coding and non-coding.

Consider case (a). The fragment can overlap with a gene in three
possible ways, with three possible shifts. If we start to read the
information one triplet after another starting from the first
letter of the fragment then we can read the gene correctly only if
the fragment overlaps it with a correct shift. In general, if the
start of the fragment is chosen randomly then we can read the gene
in three possible ways. Thus, case a) generates three possible
frequency distributions, ``shifted'' one with respect to another.

Case (b) is quite analogous and also gives three possible triplet
distributions. They are not quite independent from the ones
obtained at the step (a) for the following reason. The frequency
of triplets is in fact the same as in the case (a), the difference
is the triplets are read ``from the end to the beginning'' which
produces a kind of mirror reflection of the triplet distributions
from the case (a).

Case (c) will produce only one distribution which will be
symmetrical with respect to the ``shifts'' (or rotations) in the
first two cases, and there is a hypothesis that this is a result
of genomic sequence evolution. Let us explain it.

The vitality of a bacterium depends on the correct functioning of
all biological mechanisms. All these mechanisms are encoded in
genes, and if something wrong happens with gene sequences (for
example there is an error when DNA is duplicated), then the
organism risks becoming non-vital. Nothing is perfect in our world
and errors happen all the time, and in the DNA duplication process
as well. \index{mutations}These errors are called {\it mutations}.

The most dangerous mutations are those which change the reading
frame, i.e. letter deletions or insertions. If such a mutation
happens in the middle of a gene sequence, the rest of the gene
becomes corrupted: the reading mechanism (which reads the triplets
one by one and does not know about the mutation) will read it with
a shift. Because of this the organisms with such mutations often
die without leaving their off-spring. Conversely, if such a
mutation happens in the non-coding part, where there are no genes,
this does not lead to problems, and the organism leaves
off-spring. Thus such mutations are constantly accumulated in the
non-coding part making all three phase-specific distributions
identical. The (d) case also produces mix of triplet
distributions.

As a result, we have three distributions for case (a), three for
case (b) and one, symmetrical for the ``non-coding'' fragments
(case (c)). Because of natural statistical deviations and other
reasons we have 7 clusters of points in the multidimensional space
of triplet frequencies.

\subsection{Visualization of the 7-Cluster Structure}

\begin{figure}[t]
 \centerline{
\includegraphics[width=11.5cm]{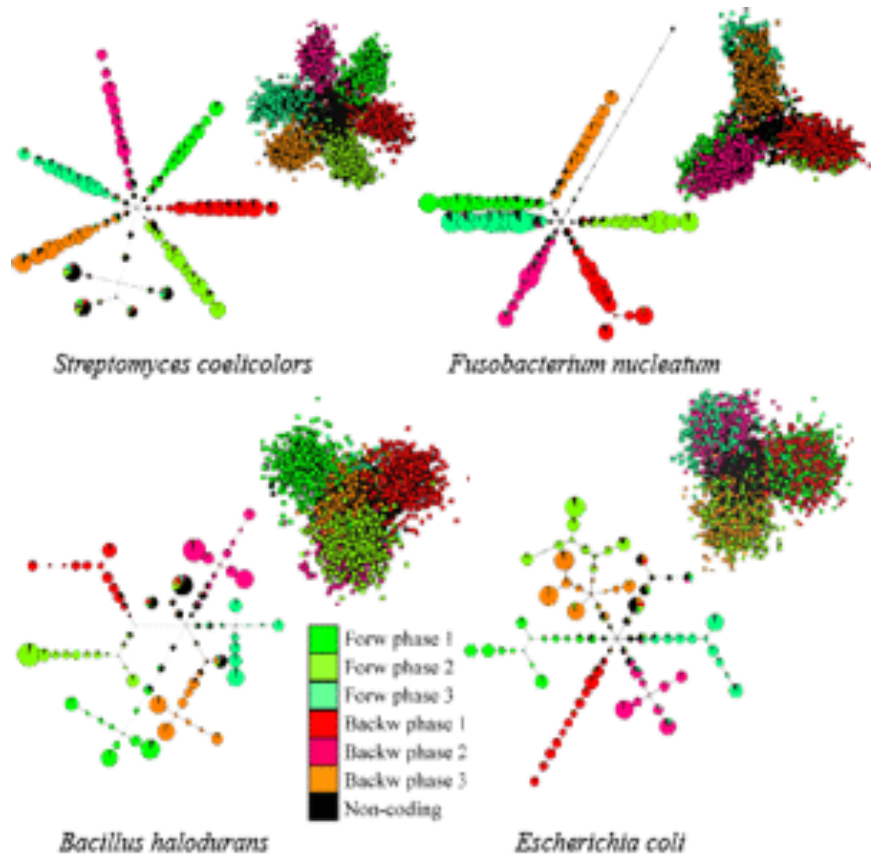}
} \caption{Seven cluster structures presented for 4 selected
genomes. A genome is represented as a collection of points (text
fragments represented by their triplet frequencies) in a
multidimensional space. Color codes correspond to 6 possible
frameshifts when a random fragment overlaps with a gene (3 in the
forward and 3 in the backward direction of the gene), and the
black color corresponds to non-coding regions. For every genome a
principal tree (``metro map'' layout) is shown together with 2D
PCA projection of the data distribution. Note that the clusters
that are mixed in the PCA plot for {\it Escherichia coli} (they
remain mixed in 3D PCA as well, see \cite{web7clusters}) are well
separated on the ``metro map''} \label{7clusters}
\end{figure}

It happens that the flower-like pattern of the 7-cluster structure
is only one of several possible \cite{7cluPhA,7clulast} when we
observe many bacterial genomes. Four ``typical'' configurations of
7-clusters observed in bacterial genomes are shown on
Fig.~\ref{7clusters}.

\index{Escherichia coli} Among these four typical configurations,
there is one called ``degenerative'' ({\it Escherichia coli} in
Fig.~\ref{7clusters}). In this configuration three clusters
corresponding to reading genes in the backward direction (reddish
clusters) overlap with three clusters corresponding to reading
genes in the forward direction (greenish clusters), when the
distribution is projected in the three-dimensional space of the
first principal components. It allows us to make a hypothesis that
the usage of triplets is symmetrical with respect to the operation
of ``complementary reversal''.

However, for a real genome of {\it Escherichia coli}, we can
observe, using the ``metro map'' representation, that the clusters
are in fact rather well separated in space. This signal is
completely hidden in the PCA plot. This is even more interesting
since we are comparing data approximation and visualization by a
one-dimensional object (principal tree) with one made by a
three-dimensional linear manifold (PCA).

\section{Visualization of Microarray Data}

\subsection{Dataset Used}

DNA microarray data is a rich source of information for molecular
biology (for a recent overview, read \cite{9Leung2003}). This
technology found numerous applications in understanding various
biological processes including cancer. It allows screening of the
expression of all genes simultaneously in a cell exposed to some
specific conditions (for example, stress, cancer, normal
conditions). Obtaining a sufficient number of observations (chips),
one can construct a table of "samples vs genes", containing
logarithms of the expression levels of typically several thousands
($n$) of genes in typically several tens ($m$) of samples.

We use data published in \cite{9Shyamsundar2005} containing gene
expression values for 10401 genes in 103 samples of normal human
tissues. The sample labels correspond to the tissue type from which
the sample was taken. This dataset was proposed for analysis for the
participants of the international workshop "Principal
manifolds-2006" which took place in Leicester, UK, in August of
2006. It can be downloaded from the workshop web-page
\cite{9PM2006Webpage}.

\subsection{Principal Tree  of Human Tissues}

\begin{figure}
 \centerline{
\includegraphics[width=8.5cm]{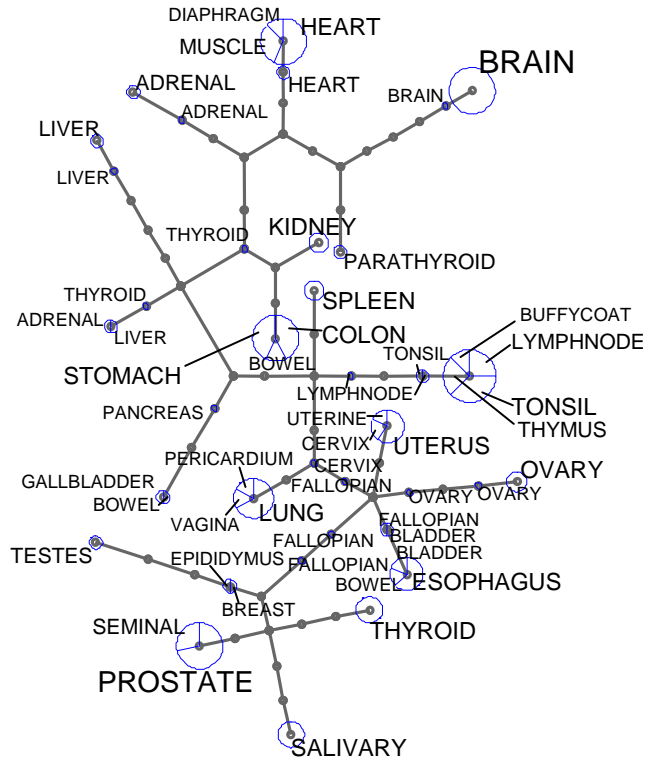}
} \caption{Principal tree of human tissues, constructed from the
gene expression microarray data. The size of the circles
corresponds to the number of points projected into this node. The
sectors show the proportion of different tissue types projected
into a node} \label{d3tree}
\end{figure}

\index{tissue} On Fig.~\ref{d3tree} a metro map representation of
the principal tree calculated for the human tissue data is shown.
To reduce the computation time we first calculated a new spatial
basis by calculating 103 linear principal components and projected
samples from the full-dimensional space into this basis. The
missing values in the dataset were treated as described in the
accompanying paper \cite{GorZin2007Springer} (a data point with
missing value(s) is represented as a line or a (hyper)plane
parallel to the corresponding coordinate axes, for which we have
missing information, and then projected into the closest point on
the linear manifold). The principal tree was then constructed
using the {\it vdaoengine} Java package available from the authors
by request. We stopped construction of the optimal principal tree
when 70 nodes were added to the tree.

One can see from the figure that most of the tissues are correctly
clustered on the tree. Moreover, tissues of similar origin are
grouped closely.

\section{Discussion}

In the continuum representation, factors are one-dimensional
continua, hence, a product of $r$ factors is represented as an
$r$-dimensional {\it cubic complex} \cite{CubMatPol} that is glued
together from $r$-dimensional parallelepipeds (``cubes"). Thus,
the factorized principal elastic graphs generate a new and, as we
can estimate now, useful construction: a principal cubic complex.
One of the obvious benefits from this construction is adaptive
dimension: the grammar approach with energy optimization develops
the necessary number of non-trivial factors, and not more. These
complexes can approximate multidimensional datasets with complex,
but still low-dimensional topology. The topology of the complex is
not prescribed, but adaptive. In that sense, they are even more
flexible than SOMs. The whole approach can be interpreted as an
intermediate between absolutely flexible {\it neural gas}
\cite{NeuralGaz} and significantly more restrictive {\it elastic
map} \cite{GorZinComp2005}. It includes as simple limit cases the
$k$-means clustering algorithm (low elasticity moduli) and
classical PCA (high $\mu$ for $S_2$ and $\mu \to \infty$ for
$S_k$, $k>2$).

We demonstrated how application of the simplest ``add a node,
bisect an edge'' grammar leads to the construction of a useful
``principal tree'' object (more precisely, branching principal
curve) which can be advantageous over the application of customary
linear PCA. Of course, more work is required to evaluate all
advantages and disadvantages which this approach gives in
comparison with existing and widely used techniques (for example,
with hierarchical clustering). However, it is clear that the
principal tree approach accompanied by the ``metro map''
representation of data, can provide additional insights into
understanding the structure of complex data distributions and can
be most suitable in some particular applications.

\end{document}